\documentclass[superscriptaddress,prl,twocolumn]{revtex4-2}

\usepackage{graphicx}
\usepackage{bm}
\usepackage{float}
\usepackage{hyperref}

\usepackage{color}

\begin{document}

\title{Micromechanics of compressive and tensile forces \\ in partially-bonded granular materials}

\author{Abrar Naseer}
\affiliation{Department of Civil Engineering, Indian Institute of Science, Bangalore, India}

\author{Karen E. Daniels}
\affiliation{Department of Physics, North Carolina State University, Raleigh, NC, USA}

\author{Tejas G. Murthy}
\affiliation{Department of Civil Engineering, Indian Institute of Science, Bangalore, India}

\date{\today}

\begin{abstract}
In granular media, the presence of even small amounts of interparticle cohesion manifests as an increase in the bulk strength and stiffness, effects that are typically associated with an increase in the average number of constraints per particle. By performing an ensemble of isotropic compression experiments, all starting from the same initial particle configuration but with varying fraction of bonded particles, we use photoelastic force measurements to identify the causes of this phenomenon at the particle-scale and meso-scale. As a function of the percentage of bonded particles, we measure a small decrease in the critical packing fraction at which jamming occurs. Above jamming, the local pressure increases predominantly for the bonded particles, measured relative to the unbonded case, through approximately equal contributions of both tensile and compressive forces. 
Histograms of the magnitude of the interparticle forces become broader for systems with more bonded particles.
We measure both pressure and coordination number as a function of distance from a bonded dimer, both of which are locally enhanced for nearest neighbors, indicating that dimers appear to act as areas of concentrated force and connectivity that improve rigidity. 

\end{abstract}
\maketitle
  
Bonding or cohering particulates is central to many industrial processes such as polymer flocculation, formation of granules in pharmaceutical products and sintering of metal particles \cite{sudhirsharma_experimental_2025}. Natural geomaterials such as limestone, sandstone, gypsiferous soils and cemented sands are formed when precipitates or organic matter bonds sand particles \cite{bernabe_effect_1992, seiphoori_formation_2020}. Adding small amounts of cohesive binders creates both additive manufacturing feedstock \cite{gilmer_additive_2021, valente_extrusion-based_2019} and infrastructure materials such as concrete and mortar \cite{de_brito_past_2021}, a technique that both dates back to Romans and is seeing new attention for creating sustainable materials \cite{goyal_physics_2021}. 
These materials differ significantly from dry granular materials because their interparticle interactions are not only compressive and frictional, but also tensile. The observed strength enhancement is reasonably well-captured in continuum models by adding a tensile strength cutoff based on empirical measurements \cite{rossi_constitutive_2024}. Partial bonding is of particular interest; it can occur as a function of progressive erosion \cite{zhang_thermodynamic-based_2022} or deformation \cite{thakur_micromechanical_2014, delenne_mechanical_2004} for which at sufficiently large strains the parent granular ensemble is recovered \cite{cuccovillo_mechanics_1999}.
Such diverse phenomena as cracking \cite{schopfer_impact_2009}, interfacial delamination in ceramic coatings \cite{ghasemi_discrete_2020}, and damage evaluation in asphalt \cite{kim_simulation_2008} have all been explained as consequences of the addition of cohesion to a particulate material.  
While there has been significant progress in understanding the mechanisms which control stiffness and rigidity in both attractive \cite{trappe_jamming_2001, martinoty_viscoelastic_2025} and frictional \cite{liu_frictional_2019,liu_spongelike_2021} jammed systems, the case of partial bonding has only begun to be investigated in spite of its ubiquity in both natural and engineered systems. Simulations address idealizations of cementation via pinned particles \cite{bhowmik_effect_2019}, dimers \cite{ozawa_creating_2023}, elongated particles \cite{mutneja_yielding_2023}, or bonding particles to form complex shapes \cite{miskin_adapting_2013}. In all of these examples, bonding provides a mechanism to improve the thermodynamic, kinetic and mechanical stability of the system, for which the enhanced mechanical properties can be attributed to the constraints placed on the degrees of freedom of the particles and the resulting improvement in structural order \cite{mutneja_yielding_2023}. 
 
In this work, we extend the use of photoelasticity --- previously used to study frictional interparticle interactions \cite{majmudar_contact_2005,daniels_photoelastic_2017,abed_zadeh_enlightening_2019} --- to permit measurements of the vector contact forces between bonded particles, allowing us to directly measure the effect of interparticle cohesion on the response of the ensemble. By considering an ensemble of isotropic compression experiments, each performed on an identical initial configuration of particles (similar to the isoconfigurational approaches of \cite{widmer-cooper_how_2004, snoeijer_force_2004}) for which only the number of bonded pairs is changed, we are able to isolate the effects of cohesion from configuration and history, both of which control the mechanical response \cite{snoeijer_force_2004, oda_initial_1972, farain_thermal_2024,dagois-bohy_soft-sphere_2012}. 
We observed, as expected, that the addition of bonded pairs shifts the jamming transition to a slightly lower packing fraction. Above jamming, the local pressure increases predominantly for the bonded dimers, creating a force distribution with a longer tail of higher force magnitudes, where tensile and compressive contact forces are equally likely to occur between the particles in the dimers. The increase in mechanical stability at the ensemble level originates from dimers acting as force sinks and improving particle connectivity, an effect that propagates to the surrounding particles as well.
\begin{figure*}
\includegraphics[width=0.9\linewidth]{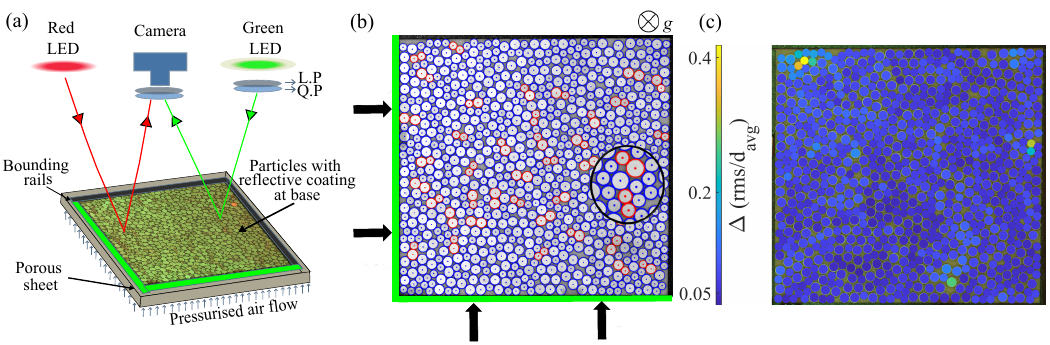}
\caption{(a) Experiment schematic: particles are confined within a square region by two motor-controlled walls (green), supported from below by airflow through a porous sheet. Monochromatic red light records all particle locations, and circularly polarized green light records stress-induced birefringence. (b) Image of the red channel, showing the blueprint particle configuration used for all experiments, with the locations of all possible bonded pairs shown by red outlines; for each run, a subset of these dimers is chosen at random. (c) A reconstruction of the particle configuration in which the color of each particle shows its root mean square centroid displacement averaged over the initial configurations of 10 realizations for the $\chi = 0\%$ case (see Supp. Mat. for histogram of $\Delta$). 
\label{fig: schematic}
}
\end{figure*}

\paragraph*{Experiments.}
We performed experiments on a quasi-two-dimensional granular ensemble of bidisperse photoelastic discs (Vishay PSM-4) floating on a porous sheet anchored on a table frame (see Fig.~\ref{fig: schematic}(a)), based on the design presented in \cite{puckett_equilibrating_2013}. The porous sheet allows the particles to float on an air cushion supplied by pressurized air, providing a near-frictionless basal boundary condition and allowing interparticle forces to dominate the measured properties. The $N = 832$ disk-shaped particles have equal populations with radii $r_1 = 5.5$~mm and $r_2 = 7.7$~mm ($\bar{r} = 6.6$~mm). They are confined by four rails within an area of $(0.4 \times 0.4)$~m$^2$; two rails impose displacement boundary conditions through a pair of linear actuators. The ensemble is biaxially compressed by equally straining it from these two orthogonal directions with a step-size of 0.38~mm, corresponding to $\Delta \phi \approx  0.001$. Each compression step is quasi-static, separated by a waiting time of 1~s.

In our experiment, we recorded the initial fabric, referred to here as the \textit{blueprint}, and re-initialize each trial to match this configuration (see Fig.~\ref{fig: schematic}(b)) by projecting an image onto the experiment and repositioning any displaced particles before beginning the next compression cycle.
We quantify the reproducibility of this blueprint fabric by calculating the RMS position for each particle across all the realizations, as shown in Fig.~\ref{fig: schematic}(c). Second, we introduce partial cohesion by randomly selecting pairs of particles from the ensemble. 
The selected pairs (dimers shown in red in Fig.~\ref{fig: schematic}(b))
are bonded together using a paraffin-based wax and then placed back to their original blueprint positions. We perform isotropic compression experiments both on a purely frictional ensemble (no bonds) as well as on partially-bonded ensembles with dimer concentration $\chi = 10\%$ (84 particles = 42 dimers), 15\%, 20\% and 25\%, with 10 realizations for each value of $\chi$.

For each compression step, a high-resolution camera (Hasselblad H4D, $3028 \times 2898$ px$^2$) captures the RGB image of the ensemble. Particles are detected via the unpolarized red channel \cite{daniels_photoelastic_2017,abed_zadeh_enlightening_2019} and each is assigned a unique identity (ID) across all the realizations, allowing  
us to track particle-level features under changes in $\chi$. 
Using light from the polarized green channel we measure the normal and tangential components ($f^n, f^t$) of the contact forces by solving the inverse problem \cite{majmudar_contact_2005,daniels_photoelastic_2017,abed_zadeh_enlightening_2019}. Our modifications to the open-source PeGS \cite{kollmer_jekollmerpegs_2024, naseer_extracting_2025} allow us to classify each bonded contact as either compressive or tensile by summing over all external forces in each dimer. 
From these vector contact forces, we calculate the stress tensor for each particle (radius $r$) by first calculating the dyadic product between each force vector $\vec f_j$  and the respective branch vector $\vec r_j$  --- followed by summation over all of its force bearing contacts (\textit{k}):
\begin{equation}
\sigma = \frac{1}{\pi r^2} \sum_{j=1}^{k} \vec{r}_j \otimes \vec{f}_J
\label{eq:stress_tensor}
\end{equation}
From the stress tensor, we calculate the particle-scale pressure as its trace.

\begin{figure}
\includegraphics[width=\linewidth]{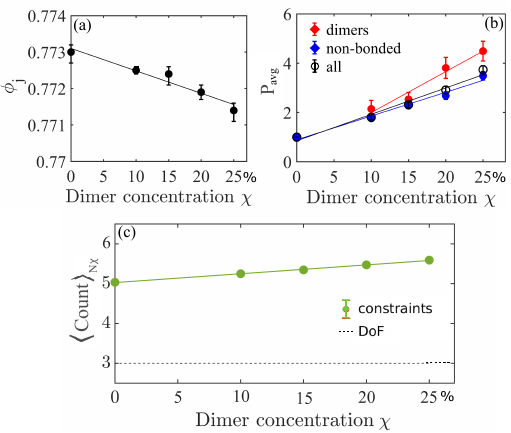}
\caption{ (a) Measured jamming packing fraction $\phi_j$ as a function of dimer concentration $\chi$, with $\phi_j$ averaged over all 10 realizations and error bars representing the error determined from the linear fit to the force response as a function of strain (see Supp. Mat). (b) Particle average of the ratio of the average single-particle pressure to the same value in the $\chi=0\%$ case. (c) While the number of degrees of freedom per particle (black dashed line) remains constant, the average number of constraints per particle (green solid line) grows, as measured from the number of contacts (considering all particles, including rattlers) for each value of $\chi$. All averages are taken over the total number of particles, $ N_{\chi}$. For each $\chi$, the normalized counts are obtained by averaging over all 10 realizations in the jammed configurations within $\phi_{rep}$.}

\label{fig: jamming}
\end{figure}

\paragraph*{Results.}
We first evaluate the effect of cohesion on the critical (jamming) packing fraction $\phi_j$. We first calculate the magnitude of the normal force per particle ($\tilde{f} = \frac{1}{N} \sum_{i=1}^{m} \left| f^n_i \right|$), where \textit{m} is the total number of force bearing contacts in the ensemble. At each $\chi$, for all the 10 realizations, we calculate $\tilde{f}$ for each strain-step, and fit best-fit lines to measure $\phi_j$, where $\phi_j$ is the point of intersection of the linear fits determining the value of $\phi$ for which the pressure rises above the background (see Supp. Mat.). As shown in Fig.~\ref{fig: jamming}(a), as $\chi$ increases, the system jams at a slightly lower packing fraction; while the difference is small, these changes are larger than the error bars for each measurement. For the remainder of the paper, we will focus our attention on the jammed states within a fixed range of packing fractions, $\phi_{rep}$ $\in$ [0.773, 0.775], which are experimentally accessible across the full range of $\chi$. Similar results were observed for other values of $\phi$. We present the averaged quantities in this paper where ${\left\langle \cdot \right\rangle_{\chi}}$ represents the average over all the 10 realizations for every $\chi$, and ${\left\langle \cdot \right\rangle_{N}}$ represents the average over all  $N$ particles in ensemble. When a quantity is computed by treating each dimer as a single particle alongside the non-bonded particles, the averaging is performed over $N_{\chi} = N(1 - \frac{\chi}{2})$ particles.

Using Eq.~\ref{eq:stress_tensor}, we evaluate pressure for each particle. For each particle (and for each $\chi$) we calculate its average pressure $\left\langle P\right\rangle_\chi$ by first averaging over all 10 realizations and then rescaling by that particle's average pressure in $\chi = 0\%$ case (see Supp. Mat.). Fig.~\ref{fig: jamming}(b) presents the values obtained by averaging these rescaled values over all $N$ particles: $ P_{avg} = \left\langle \frac{\left\langle P \right\rangle_{\chi}}{\left\langle P \right\rangle_{\chi = 0}} \right\rangle_N$ for each $\chi$. Dimers (red markers) show a higher increase in pressure values compared to the non-bonded particles (blue markers), indicating that forces are preferentially carried within the dimers. This effect is significant: at $\chi=10\%$, the pressure in dimers increases by a factor of 2, and for $\chi=25\%$, the increase is four-fold. By considering the average pressure over all particles (black open circles), and noting that each is measured for the same wall-displacement (fixed $\phi_{rep}$), we can additionally observe that these measurements correspond to an increased elastic modulus (stiffness) as a function of $\chi$. 

There are two ways to interpret these observations. First, we note that the non-convex geometry of the dimers creates the opportunity for interlocking between particles, which on its own could enhance the ensemble strength \cite{gravish_entangled_2012,athanassiadis_particle_2014}. Second, the excess in the number of constraints as compared to the degrees of freedom (DoF) is a hallmark of rigidity \cite{liu_spongelike_2021}. Figure~\ref {fig: jamming}(c) shows both quantities averaged over $ N_{\chi}$, with dimers acting as a single particle having two translational and one rotational DoF. The number of constraints is measured from the number of contacts (considering all particles, including rattlers) for each value of $\chi$, while the number of degrees of freedom is fixed at 3 (two translational and one rotational).  

\begin{figure}
\includegraphics[width=\linewidth]{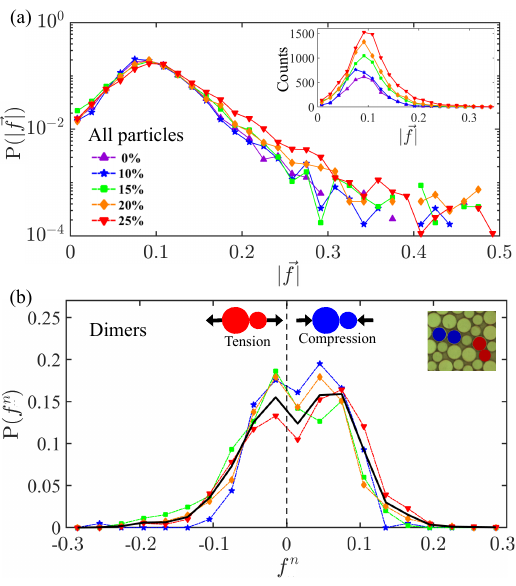}
\caption{ 
(a) Probability density function of contact force magnitudes $|{\vec f}|$ measured for all particles at fixed $\chi$ and fixed $\phi_{rep}$. Inset: raw number of counts for the same, showing the increase in total number of contacts as a function of $\chi$.  (b) Probability density function of normal force component $f^n$ measured for all bonded contacts at fixed $\chi$ and fixed $\phi_{rep}$, with negative values tensile and positive values compressive. The black line shows the average over all 4 values of $\chi$. Inset: subset of particles from $\chi = 20\%$, where colored particles are the dimers. The red ones have a higher frequency of tensile forces transmitting through their interparticle bond than the compressive forces, whereas the blue ones have a higher occurrence of compressive forces, measured across $\phi_{rep}$.}
\label{fig: fdist}
\end{figure}

Fig.~\ref{fig: fdist}(a) shows the probability distribution of the magnitude of the contact force ($|{\vec f}|$) for all particles, both non-bonded and dimers. All of these histograms show the usual peak near the mean force, followed by an exponential decay at higher forces. In systems with more dimers (higher $\chi$), the peak of the distribution is at a higher $|{\vec f}|$, the distribution has a longer tail, and the total number of contact forces increases (see inset). All of these indicate a stronger force network covering a larger fraction of the particles under increased levels of bonding. Within the bonded dimers, these increased forces arise from both compressional and tensile forces. Fig.~\ref{fig: fdist}(b) shows histograms of contact normal forces ($f^n$) within bonds connecting the two particles in each dimer,  with positive $f^n$ indicating compression and negative $f^n$ indicating tension. For all $\chi$, the distribution exhibits a bimodal distribution, with the compressive peak at a somewhat higher force than the tension peak, without any strong $\chi$-dependence. The inset illustrates the relative occurrence of both tensile and compressive forces through the bonds (see Supp. Mat. for details). By being able to detangle the sign of the forces, we ascertain the existence of tensile interactions, a phenomenon that has been hypothesized for previous simulations \cite{estrada_simulation_2010,heinze_stress_2020} studying the micromechanics of cohesive granular materials. 

\begin{figure}
\includegraphics[width=\linewidth]{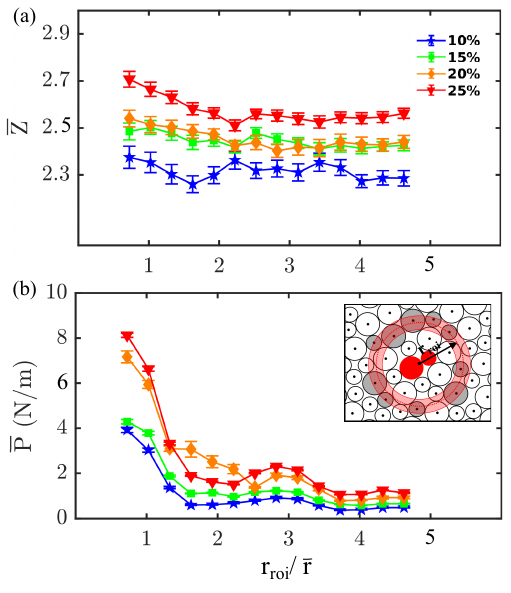}
\caption{Measured average (a)  coordination number $\bar{Z}$ and (b) pressure $\bar{P}$ of all particles within a ring-shaped region of interest (RoI) of varying radius and constant thickness equal to the average particle radius, $\bar{r}$ (as shown in the inset of b). For pressure, the averages represented by the overbar are obtained within the RoI, using the equivalent of Eq.~\ref{eq:stress_tensor} but summing only over those particles whose centers are located within the RoI. For each $\chi$, these parameters are obtained by averaging over all 10 realizations in the jammed configurations within $\phi_{rep}$.
}
\label{fig: roi}
\end{figure}

To examine the length scale over which the bonding influences the mesoscale, we measure the average coordination number $\bar{Z}$ and pressure $\bar{P}$ as a function of distance from each dimer. We construct a ring-shaped region of interest (RoI, see inset in Fig.~\ref{fig: roi}) around each dimer, and calculate the averages over all particles whose centroids are located within that ring, shown in gray. As shown in Fig.~\ref {fig: roi}, all values of $\chi$ show higher $\bar{Z}$ and $\bar{P}$ in the vicinity of a dimer, an effect which decays past the nearest-neighbor interactions. At a fixed RoI radius, systems with higher $\chi$ experience both a higher coordination number and larger pressure. In this sense, dimers appear to act as areas of concentrated force and connectivity,  which is reflected by an increase in mechanical stability at the ensemble level.

\paragraph{Discussion and Conclusions.} 
We perform experiments with a single blueprint fabric (similar to the isoconfigurational ensemble), in order to separate the effects of interparticle cohesion without the confounding effects of variability due to particular configurations \cite{dagois-bohy_soft-sphere_2012}. This method allows us to directly examine the effects of small tensile forces on the jamming and stiffness of partially-bonded particles. We find that even this small population of tensile forces enhances the stability of the force network, resulting in macroscopic behaviors such as higher stiffness and earlier jamming. Bonded particles act as areas of increased coordination number and pressure, effects which extend to a mesoscale (beyond the nearest-neighbors).  These improvements in ensemble-level behavior are approximately linear as a function of dimer concentration, which are kept small enough that the network of dimers do not themselves percolate their regions of influence. Adapting the techniques of rigidity percolation \cite{liu_spongelike_2021} to address bonded particles, via frozen constraints, would be useful for directly addressing the development of rigidity under these conditions. 

These findings are also consistent with previous results from ensemble-level experiments and simulations \cite{delenne_mechanical_2004,li_experimental_2015,heinze_stress_2020,estrada_simulation_2010}, reinforcing the earlier understanding of the presence of tensile forces in cohesive granular ensembles and their overall micromechanics. The study highlights the impact of cohesion on the force landscape, providing a basis for deeper exploration into the modeling of force networks \cite{tordesillas_network_2015}. In addition to providing insights into the micromechanics of cohesive granular materials, our study offers an experimental underpinning to creating highly stable equilibrium glasses by bonding monomers to form dimers \cite{ozawa_creating_2023}.

\paragraph{Acknowledgments.} 
We are grateful to funding from the National Science Foundation (DMR-2104986), Anusandhan National Research Foundation (ANRF-CRG/2022/003750)  and the Fulbright-Nehru fellowship program.

\bibliography{grain-cement}

\end{document}


\title{Supplementary Material: Micromechanics of compressive and tensile forces in partially-bonded granular materials}

\author{Abrar Naseer}
\affiliation{Department of Civil Engineering, Indian Institute of Science, Bangalore, India}

\author{Karen E. Daniels}
\affiliation{Department of Physics, North Carolina State University, Raleigh, NC, USA}

\author{Tejas G. Murthy}
\affiliation{Department of Civil Engineering, Indian Institute of Science, Bangalore, India}

\date{\today}

\maketitle

\appendix
\section{Blueprint fabric}
\begin{figure}[h]
\includegraphics{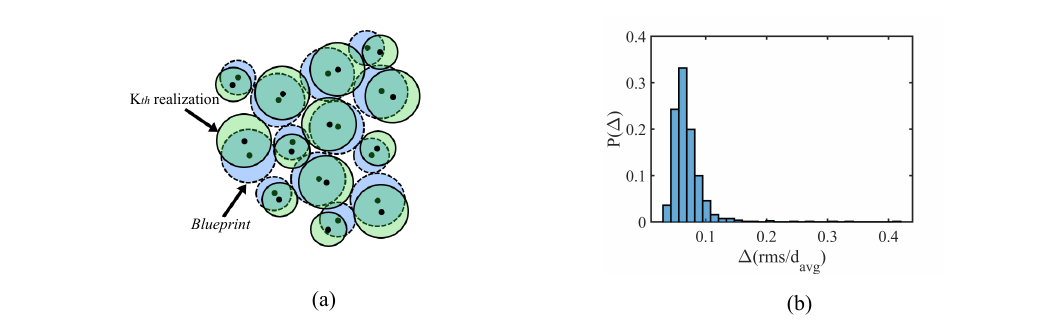}
\caption{\label{fig:blueprint} Preserving the initial fabric: (a) Schematic of an initial configuration (green solid circles) overlaid with the \textit{blueprint} (blue dotted circles). (b) Probability distribution of these RMS values across all 10 realizations, from Fig. 1(c) of the main paper.}
\end{figure}

A key technique of this paper is to preserve the initial fabric in order to detangle its effect from cohesion. This was achieved by manually placing particles at predefined locations in 2D space. The very first trial that we conducted in our suite of experiments had a random initial fabric. This fabric was then replicated in all the subsequent trials and hence forms the \textit{blueprint}. In the subsequent realizations, each particle was placed at the location where it was in the blueprint. The schematic in Fig. \ref{fig:blueprint}(a) illustrates this, a version that exaggerates the deviation of particles (green) from their expected position (blue). The blueprint is stored as an image by doing particle detection using a Hough transform implementation in MATLAB. An image projector positioned above the air table projects the image, showing tracked centroids and particle shape outlines, onto the space between the bounding rails (refer to the experimental setup schematic in Fig. 1(a) of the main paper). All $N = 832$ particles were then carefully placed such that the projected outlines of the expected position coincide with the particle locations. To quantify the matching of newly constructed fabric with the blueprint, we calculate the root mean square (RMS) values of the particle centroids. The color-coded image (Fig. 1(c) of main paper) shows these values for each particle which is less than the 10\% of the average particle diameter. The histogram (Fig. \ref{fig:blueprint}(c)) shows the distribution of the RMS values for all 10 realizations conducted for $0\%$ ensemble.

\clearpage 

\section{Jamming packing fraction}
\begin{figure}[h]
\includegraphics[width=\linewidth]{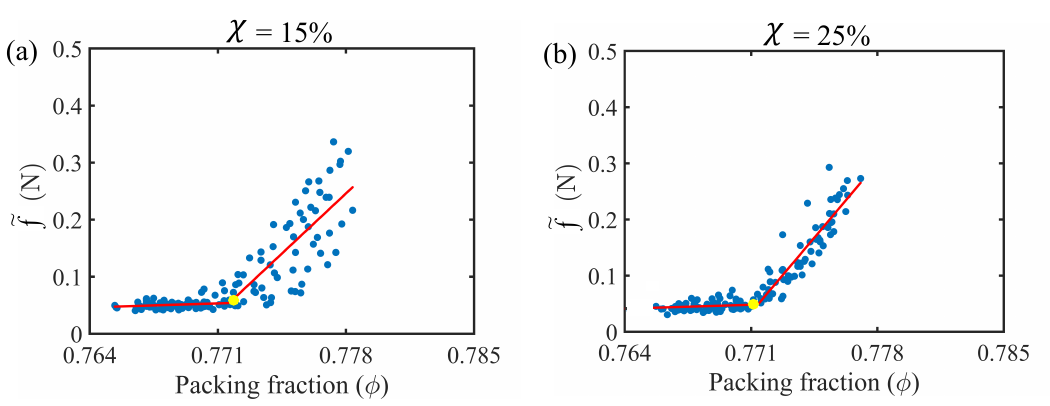}
\caption{\label{fig:jamming} 
Average of the magnitudes of the normal contact forces acting on particles ($\tilde{f}$) as a function of packing fraction. (a) $\chi = 15\%$ and (b) $\chi =25\%$ best-fit lines (red) fitted over all the strain-steps (blue markers) in 10 realizations. The yellow point marks the jamming packing fraction ($\phi_j$) of the respective ensemble.}
\end{figure}

For each strain-step we evaluate the average magnitude of normal contact forces acting per particle $$\tilde{f} = \frac{1}{N} \sum_{i=1}^{m} \left| f^n_i \right|$$ where $f^n$ is the normal contact force, $N$ is the number of particles, \textit{m} is the total number of force bearing contacts. The evolution of $\tilde{f}$ is approximately constant at low $\phi$ (representing background noise) and shows an increase as the ensemble compresses to higher $\phi$. The ensemble starts to resist the external force and builds mechanical resistance in response to the external perturbation, which is seen as an increase in $\tilde{f}$. This increase marks the jamming transition of the ensemble, where the transition point ($\phi_j$) is the point of intersection of the two best-fit lines (red) for each $\chi$. Comparing (a) and (b), with the increase in $\chi$, the ensemble shows slightly lower value of $\phi_j$. The decrease in $\phi_j$ with increase in $\chi$ is plotted in main paper, Fig 2(a).

\clearpage 

\section{Pressure at particle-level}
\begin{figure}[h]
\includegraphics{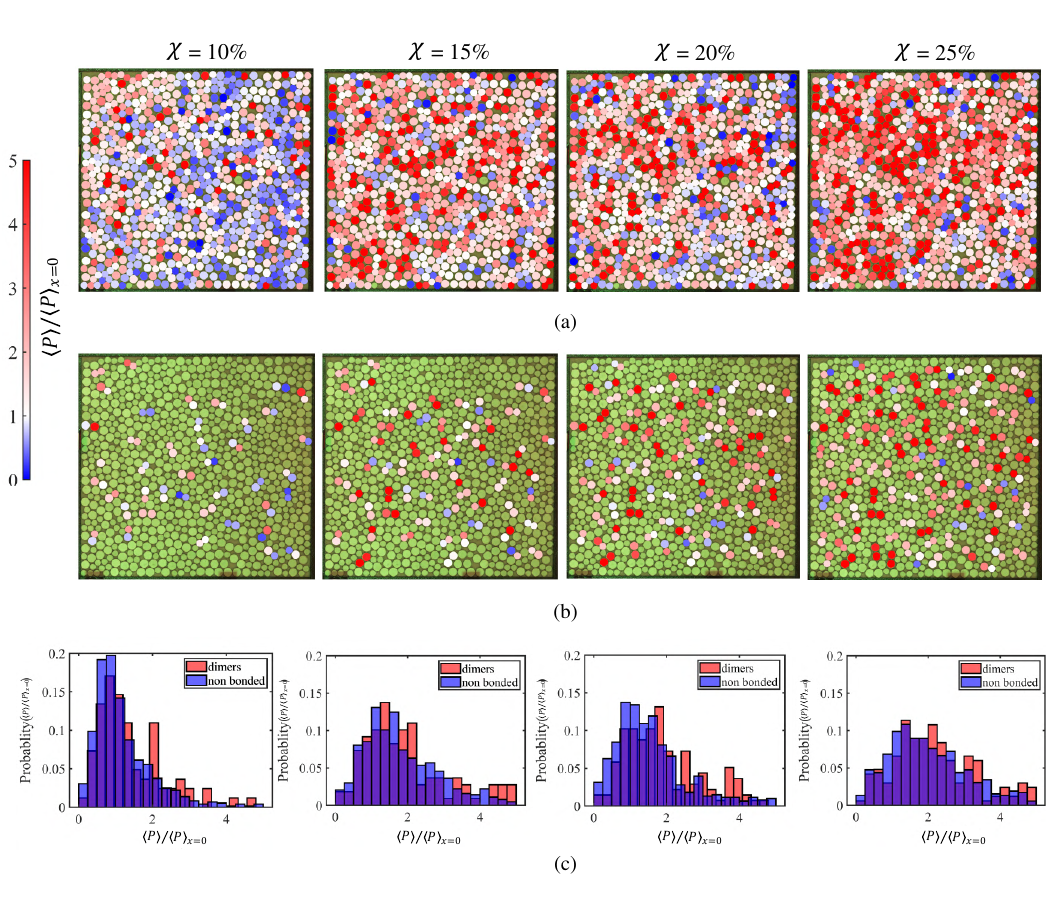}
\caption{\label{fig:hist} (a) Colorbar showing the pressure acting on each particle in a cohesive ensemble in comparison to 0\% bonded case. For each particle the pressure is averaged over all 10 realizations and then this value is scaled by the particle's average pressure for the $\chi$ = 0 case. (b) Panel showing the version of (a), with only dimers color labelled according to their scaled pressure values. (c) Comparison of probability distribution of these values for non-bonded particles and dimers.}
\end{figure}

We measure the pressure in each particle by calculating the trace of the Cauchy stress tensor defined by Eq. 1 in the main paper. For each particle the values are averaged over all 10 trials and scaled by its value for the $\chi = 0$ case ($\frac{\left\langle P \right\rangle_{\chi}}{\left\langle P \right\rangle_{\chi = 0}}$). It can been seen in Fig. \ref{fig:hist}(a) the population of particles bearing higher pressures increases with $\chi$ (particles colored in red). Panel (c) shows that the effect is more dominant in dimers, as most of them appear in red. To quantify this we plot the probability distribution of pressure in non-bonded particles and dimers. As the dimer concentration increases, the distribution shifts rightward for both non-bonded particles and dimers, implying a greater propensity of higher contact forces in the ensemble.

\clearpage 

\section{Nature of forces transmitting through the bonds}
\begin{figure}[h]
\includegraphics[width=0.70\linewidth]{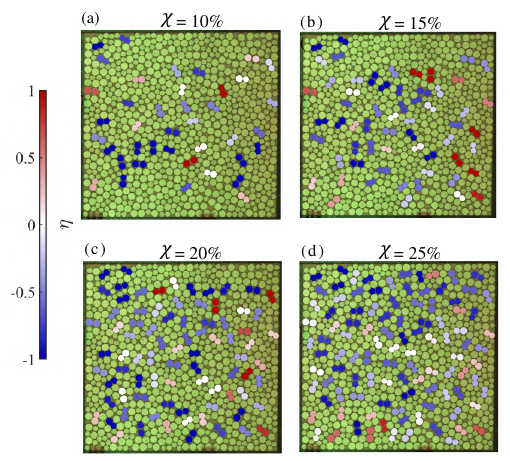}
\caption{\label{fig:ratio} Relative frequency of tensile forces vs compressive forces
}
\end{figure}

We evaluate the nature of the force transmitting through the bonds. The parameter \(\eta\) is defined as ratio of difference of number of tensile forces and compressive forces with respect to the total number of forces ($\eta = \frac{n^+ - n^-}{n}$) that propagate through the bonds averaged over all the strain-steps within $\phi_{rep}$. The quantity $\eta$ is positive if the occurrence of tensile forces is larger than the compressive forces (dimers in red) while as it is negative if compressive forces have a higher occurrence than the tensile forces (dimers in blue). The tensile bearing capacity induced by bonds results in a bimodal nature of force distribution (plotted in the main paper, Fig. 3(b)).  

\clearpage 